\documentclass[11pt,twoside]{article}

\usepackage{asp2006}
\usepackage{epsf}
\usepackage{psfig}
\usepackage{lscape,graphicx}

\begin{document}
\title{Thioformaldehyde Emission from the Massive Star-Forming Region DR21(OH)}   
\author{Alwyn Wootten and Jeffrey Mangum}   
\affil{NRAO, 520 Edgemont Road, Charlottesville, Virginia 22903}    

\begin{abstract} 
Using arguments parallel to those used in support of using H$_2$CO as a sensitive probe of temperature and density in molecular clouds, we measured the J=7$\rightarrow$6 and J=10$\rightarrow$9 transitions of thioformaldehyde (H$_2$CS) in several hot core sources.  The goal here was to investigate more closely the conditions giving rise to H$_2$CS emission in cloud cores containing young stars by modelling several transitions.  The H$_2$CS molecule is a slightly asymmetric rotor, a heavier analogue to H$_2$CO.  As in H$_2$CO, transitions occur closely spaced in frequency, though they are substantially separated in energy.  Transitions of H$_2$CS originating from the K=0, 1, 2, 3, and 4 ladders in the 230 and 345 GHz windows can productively be used to constrain densities and temperatures.  As a first step in developing the use of these transitions as thermometers and densitometers, we surveyed and modeled the emission from well known warm dense cores.
\end{abstract}


\section{Introduction}   
The H$_2$CS molecule is asymmetric and planar as is the case for H$_2$CO; it has C$_{2v}$ symmetry. The molecular asymmetry results in a substantial dipole moment of 1.65 Debye. 
An advantage of the spectrum of a slightly asymmetric rotor molecule is that lines arising between levels of different energies lie nearby in the spectrum, affording a gauge of excitation while observing with a single or only slightly different tuning on one telescope during one session.  Calibration uncertainties are thereby minimized.  In a single spectral region the lines of both formaldehyde and thioformaldehyde occur between two states with the same angular momentum quantum number, J.  As there may be numerous values for the projection of the angular momentum vector on the main axis of the molecule, there are many rotational states with a single angular momentum quantum number J; the quantum number denoting these projections are usually denoted K.
Except for the case that K=0, the spins of the hydrogen atoms further split each level denoted by a particular value of K into two levels.  The spins of the hydrogen atoms also segregate two states of the molecule, ortho states in which the value of K is odd, and para states in which the value of K is even.  As a result there are often an abundance of lines in any broad frequency window which arise from many states of excitation.  Measuring these lines can result in a measure of the state of excitation of the gas in which they arise, in a similar manner to that for which H$_2$CO lines are used (Mangum and Wootten 1993).
\section{Observations}    
Measurements of a variety of transitions in the 230 and 345 GHz
atmospheric windows were made on 2006/04/29--05/04 using the facility receivers of the Caltech Submillimeter Observatory (CSO).  Pointing was checked several times each night and was found to be stable to within 5\arcsec.  Antenna efficiency was checked using position-switched measurements of Jupiter ($\Theta$=44.6Ó,  T$_{jupiter}\sim$ 170 K).  At 240 and 345 GHz the CSO main beam FWHM measures $\sim$30\arcsec and $\sim$21\arcsec respectively.  Using the formalism described in Mangum (1993) we derive $\eta_{mb}$ $\sim$ 0.71 and $\sim$0.61 at 240 and 345 GHz, respectively; results are given in Table 1.  Lines observed in DR21(OH) are illustrated (Fig. 1); line ratios measured toward the source and derived parameters are given in Table 2. 

\begin{figure}[t]
\begin{center}
\includegraphics[width=2.3in] {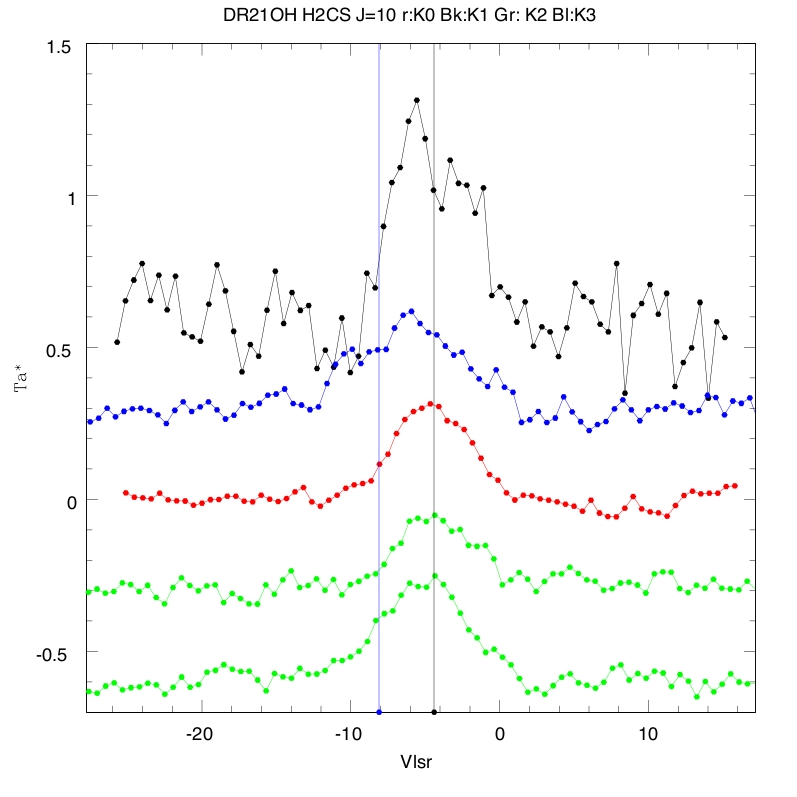} 
\caption{DR21(OH) spectra of the H$_2$CS lines at 342.9 GHz (K=0 center) , 343.8 GHz and 343.3 (K=2, upper and lower respectively of the bottom lines); and 343.5 GHz (K=1, upper) and 343.4 GHz (K=3, second from top; the left vertical flag marks the location of the second component of the blend).  The right vertical flag marks the adopted frequencies (from Splatalogue/SLAIM). 
  }
\label{fig1}
\end{center}
\end{figure}

The lines in Table 1 cluster in frequency,
 allowing a single suitable bandwidth receiver to measure them all simultaneously or with little retuning (hence with a more accurate calibration).  However, their energies are substantially different.  In one observation per band, then, one can observe several line ratios.  As H$_2$CS has a lower abundance than H$_2$CO, and because there are more energy levels in any given interval, the molecular populations are spread thinly and the emission is expected to be optically thin.  Line ratios should, therefore, provide a fairly robust estimate of excitation.  For most sources in our survey, we have observed several lines at 1.3mm and several additional lines at .85mm.  For a convenient excitation measure we employ line ratios, for the 1.3mm para-H$_2$CS lines we define R$^{J7}_{02}$ as the 7$_{07}$--6$_{06}$ to 7$_{26,5}$  -- 6$_{25,4}$ line ratio, given in Table 2 (where we have averaged the K=2 lines); a similar quantity for the .85mm para-H$_2$CS lines,  R$^{J10}_{02}$ = T$_R^*$(10$_{0,10}$  -- 9$_{0,9}$)/ T$_R^*$(10$_{2,8}$  -- 9$_{2,7}$), where again the K=2 lines are averaged,  and for the 0.85mm ortho-H$_2$CS lines R$^{J10}_{13}$ = T$_R^*$(10$_{1,9}$  -- 9$_{1,8}$)/ T$_R^*$(10$_{3,7}$  -- 9$_{3,6}$).  
\begin{table}[ht]	
\caption{H$_2$CS Results: DR21(OH) J2000  20:39:01.1  42:22:48}
\begin{center}
\label{tab:results}
%
%
%

\begin{tabular}{llllll}
\hline
 Transition & Frequency&T$^*_A$ & V$_{LSR}$ & $\Delta$V & T$^{mod}_R$ \\
& (GHz) & (K) &(km/s) & (km/s) & (K) \\
\hline\hline
 $10_{0,10}\rightarrow9_{0,9}$ &342.946456& $0.32\pm0.01$ & $-4.5$ & 5.8 & 0.30 \\
 $10_{28}\rightarrow9_{27}$ &343.321887& $0.24\pm0.02$ & $-4.4$ & 6.1 & 0.07 \\
  $10_{3,8/7}\rightarrow9_{3,7/6}$ &343.413647& $0.17\pm0.01$ & $-4.4$ & 5.2 & ... \\
 $10_{29}\rightarrow9_{28}$ &343.812941& $0.33\pm0.02$ & $-5.1$ & 6.9  & 0.07\\
 $10_{19}\rightarrow9_{18}$ &348.53424& $0.58\pm0.10$ & $-5.0$ & 6.5 & 0.61\\
 $7_{07}\rightarrow6_{06}$ &240.26632& $0.49\pm0.02$ & $-3.2$ & 5.9 & 0.51 \\
 $7_{26}\rightarrow6_{25}$ &240.38175& $0.12\pm0.03$ & $-4.0$ & 7.1 & 0.13 \\
 $7_{3,5/4}\rightarrow6_{3,4/3}$ &240.39281& $0.15\pm0.03$ & $-4.0$ & 5.0 & 0.11 \\
 $7_{25}\rightarrow6_{24}$ &240.39348& $0.13\pm0.02$ & $-3.8$ & 5.1 & 0.13 \\
\hline
\end{tabular}
\end{center}

\end{table}
\begin{table}[ht]	
\caption{H$_2$CS Line Ratios: DR21(OH) }
\begin{center}
\label{tab:ratios}
\begin{tabular}{crrr}
\hline
 Ratio & Value & T$_k$  & N(H$_2$)\\
 & & (K) & cm$^{-3}$ \\
\hline\hline
$7_{0,7}\rightarrow6_{0,6}$/$7_{2,6/5}\rightarrow6_{2,5/4}$ & 3.93 $\pm$ 0.82& 35$^{36}_{33}$ & Note a\\
 $10_{0,10}\rightarrow9_{0,9}$/$10_{2,8/9}\rightarrow9_{2,7/8}$ & 1.15 $\pm$ 0.14 & 23$^{28}_{19}$ &  Note a\\
 $10_{1,9}\rightarrow9_{1,8}$/$10_{3,8/7}\rightarrow9_{3,7/6}$ & 3.42 $\pm$ 0.62 & 58$^{67}_{49}$  & Note a \\
 \hline
$7_{0,7}\rightarrow6_{0,6}$/$10_{0,10}\rightarrow9_{0,9}$ &1.53 $\pm$ 0.08& Note b & 5 $\times$ 10$^7$ \\
$7_{2,6/5}\rightarrow6_{2,5/4}$/$10_{2,8/9}\rightarrow9_{2,7/8}$ & 0.44 $\pm$ 0.14& Note b & \\
$7_{3,5/4}\rightarrow6_{3,4/3}$/$10_{3,8/7}\rightarrow9_{3,7/6}$ & 0.88 $\pm$ 0.18& Note b & \\
\hline
\end{tabular}

\end{center}
a.  These ratios measure temperature more effectively than density; n(H$_2$)=5 $\times$ 10$^7$ cm$^{-3}$ assumed.
b.  These ratios measure density more effectively than temperature; T$_k$=37K assumed..
\end{table}
%
%
%

\section{Interpretation} 

We have used a spherical large velocity gradient (LVG) treatment of radiative transfer (RADEX; van der Tak et al. 2007) to model the excitation of H$_2$CS within the context of a single temperature and density model of the core region of DR21(OH) in which the lines we have observed arise.  In this simple model, the cloud is constrained to be a very simple spherical entity with uniform spatial density, kinetic temperature and molecular column density.  The transfer of radiation is assumed to be governed by these local conditions.  The solutions to a statistical equilibrium model of the molecule and the transfer of radiation can be expressed in terms of the radiation temperatures of the molecular transitions and their associated optical depths.  The observed spectrum is then compared to the model spectrum.  We have included 71 transitions between 34 levels with energies below 112 cm$^{-1}$ of para-H$_2$CS and 106 transitions between 42 levels of ortho-H$_2$CS in our models.

The collisions of the molecule are a key to modeling statistical equilibrium, as collisions provide the link in the model to the local spatial density of collision partners, primarily H$_2$.  Cross sections for the excitation of thioformaldehyde by collision with molecular hydrogen are unavailable.  We have used a scaled version of the collisional rates for formaldehyde with He calculated by Green (1991).  In accord with previous suggestions, we have scaled the rates by a factor of 2.2 to approximate the increased interaction with H$_2$ as collision partners.  A similar treatment has been applied to observations of other massive cores for H$_2$CS by van der Tak et al. (2003).

From the three measured line ratios in the second table, between K=0 and K=2 lines at 240 and 340 GHz, and between K=1 and K=3 lines at 340 GHz, we find a weighted mean average kinetic temperature of 37$\pm$7K.  From the $7_{0,7}\rightarrow6_{0,6}$/$10_{0,10}\rightarrow9_{0,9}$ (E$_u$=91K) excitation ratios, we find a mean density of n(H$_2$)=$5 \times 10^7 cm^{=3}$.  The total column density of H$_2$CS is N(H$_2$CS)=$8.8 \times 10^{13} cm^{-3}$, for the measured linewidth of 5 km s$^{-1}$.  The ortho to para ratio is three. Using these parameters, we calculate the line intensities written in Table 1.  The agreement with the observations is excellent, with the exception of the  $10_{2,9/8}\rightarrow9_{2,8/7}$ lines, which are too weak by a significant factor.  The latter are the higher excitation lines in one of the two high excitation density measures available in our data--the $7_{2,6/5}\rightarrow6_{2,5/4}$/$10_{2,8/9}\rightarrow9_{2,7/8}$ (E$_u$=143K) and $7_{3,5/4}\rightarrow6_{3,4/3}$/$10_{3,8/7}\rightarrow9_{3,7/6}$ (E$_u$=209K) ratios.  Unfortunately,  the present cross sections we have employed do not allow a prediction of the $10_{3,8/7}\rightarrow9_{3,7/6}$ line.  However, we note that while the ratio of the K=0 lines is high--the lower frequency line is much stronger than the higher frequency line--the opposite is true of the higher excitation ratios, where it is the higher excitation line which is strongest.  The beamsize with which the higher frequency lines are measured is smaller than that with which the lower frequency lines was measured by a factor of 0.7.  If the high excitation emission were to arise in a source smaller than the beam, one would expect a relatively stronger intensity from the higher frequency transition.  

We conclude that H$_2$CS provides a useful tool to measure the temperature of warm regions, though some details remain to be explored.  CH$_3$CN is often used for a temperature probe, but as it is formed in a gas phase chemistry only when high temperatures occur, it can be a biased probe. H$_2$CS is formed in an exactly analogous fashion to H$_2$CO in a chemistry that is fairly temperature-insensitive below 300K.  It offers advantages over H$_2$CO in the large number of available lines, and the much lower abundance (down by more than an order of magnitude).  The large number of lines, available to wideband spectrometers such as that provided by ALMA, will allow us to deconstruct temperature/density along a line of sight.

\acknowledgements 
We thank E. van Dishoeck for providing the collisional rates for H$_2$CS.  The
  Caltech Submillimeter Observatory is supported by the National
  Science Foundation under award AST-0540882.


\end{document}